\documentclass[aps,prl,twocolumn]{revtex4}

\usepackage{graphicx}
\usepackage{dcolumn}
\usepackage[latin1]{inputenc}
\usepackage{ifthen}

\usepackage{natbib}

\usepackage{hyperref}

\newcommand{\ket}[1]{\left| #1 \right\rangle}

\begin{document}

\title{Topologically decoherence-protected qubits with trapped ions}
\author{P.Milman,$^{1}$ , W. Maineult$^{1}$, S. Guibal$^{1}$, L. Guidoni $^1$%
, B. Douçot $^2$, L. Ioffe $^3$, T. Coudreau $^{1}$}
\email{coudreau@spectro.jussieu.fr}
\affiliation{$^{1}$ Laboratoire Matériaux et Phénomènes Quantiques, CNRS UMR 7162,
Université Denis Diderot, 2 Place Jussieu, 75005 Paris, France\\
$^{2}$ Laboratoire de Physique Théorique et Hautes Energies, CNRS UMR 7589,
Universités Paris 6 et 7, 4, place Jussieu, 75005 Paris, France\\
$^3$ Department of Physics and Astronomy, Center for Materials Theory,
Rutgers University, 136 Frelinghuysen Road, Piscataway, New Jersey 08854, USA}
\date{\today}

\begin{abstract}
We show that trapped ions can be used to simulate a highly symmetrical
Hamiltonian with eingenstates naturally protected against local sources of
decoherence. This Hamiltonian involves long range coupling between particles
and provides a more efficient protection than nearest neighbor models
discussed in previous works. Our results open the perspective of
experimentally realizing in controlled atomic systems, complex entangled
states with decoherence times up to nine orders of magnitude longer than
isolated quantum systems.
\end{abstract}

\maketitle

It is universally accepted that quantum computation would be able to solve
diverse classes of hard problems more efficiently than its classical
counterpart \cite{Shor1994,Grover1997}. However, its practical realization
is made difficult by the conflicting requirements imposed by the absence of
decoherence, qubit manipulation and scaling \cite{Joos1996}. Generally,
larger dimensional systems are in general more sensitive to decoherence, so
that scaling of quantum information processors is a huge experimental
challenge. In the present work, we show how to avoid decoherence and protect
complex multiparticle quantum states (that can define a qubit) from its
effects in an experimentally realizable trapped ion system.

There are two conceptually different approaches to suppress the decoherence
to a level required for quantum error correction. One is to reduce directly
the physical noise that leads to decoherence. It is however very difficult
due to the stringent requirements imposed by quantum error correction~\cite%
{Gottesman}. Another is to encode qubits in decoherence free subspaces
(DFS) which are decoupled (at least to first order) to the dominant
sources of noise. The efficiency of the latter strategy is
demonstrated by experiments showing a dramatic increase in the
decoherence times in diverse physical systems: atom clouds
\cite{Kimble2005}, trapped ions \cite{Roos2004} or superconducting
circuits \cite{cooper} in which noise mostly originates in the
fluctuating electric or magnetic fields. In trapped ion systems, DFS
are usually hyperfine states that can either be symmetrically coupled
to the magnetic field \cite{Wine2005}, suppressing dephasing to first
order, or have the same energy \cite{Haff2005}, avoiding spontaneous
emission.

Here we propose a design that provides higher degree decoupling from
environment. The main
idea~\cite{Kitaev1997,Denis2002,Ioffe2002,Doucot2005,Dorier2005,Zoller2006,Lukin2007} of our approach is to implement a long range
Hamiltonian which non-local symmetries ensure that not only the linear
but also quadratic, cubic, etc coupling to the environment of the
lowest doublet vanishes, providing thereby a high degree of protection
against the effects of most sources of physical noise. The Hamiltonian
introduced here is more efficient for decoherence protection than
previously studied ones, as is shown below. In our scheme, the
protected states are complex entangled states of $N^{2}$ spin $1/2$
like particles. It is an example of multidimensional quantum state
that is far more robust against decoherence than its individual
elements.

In the second part of this letter we propose the implementation of the
model Hamiltonian in trapped ion systems. The long range interaction
responsible for protection in our Hamiltonian is particularly adapted
to trapped ions systems. We show how this type of experimental system
that has been already employed to implement basic operations of
quantum logics~\cite{qubitwineland} can be also used to realize the
protection by topological order.

We start by introducing the long range Hamiltonian acting in a collection of
two level systems.
\begin{equation}
H=-J_{x}\sum_{i}^{N}\left( \sum_{j}^{N}\sigma _{i,j}^{x}\right)
^{2}-J_{y}\sum_{j}^{N}\left( \sum_{i}^{N}\sigma _{i,j}^{y}\right) ^{2}.
\label{eq:infinite}
\end{equation}%
Here we label each particle by its position in a two dimensional
lattice: $\sigma _{i,j}^{x,y}$ are Pauli matrices of the spin situated
at the intersection of the $i^{th}$ row and $j^{th}$ column. This
Hamiltonian couples the particles in the same row by $\sigma
_{i,j}^{x}\sigma _{i,k}^{x}$ interaction and those in the same column
by $\sigma _{i,k}^{y}\sigma _{j,k}^{y}$ interaction. Consider the
projections of all spins in a given row onto $y$-axis. Evidently the
projection is preserved by the column interaction while row
interaction might change it by 0 or 2 units. Thus all terms in the
Hamiltonian preserve the parity of the $y-$projection of the spins in
one row. Mathematically, it can be described by non-local symmetry
transformations generated by the operators $P_{i}=\prod_{j}\sigma
_{i,j}^{y}$ and $Q_{j}=\prod_{i}\sigma _{i,j}^{x}$ that involve
product of all spins in each row or column. It can be easily shown
that the operators of a given set commute with all others in the same
set, \emph{i.e.}  $[P_{i},P_{j}]=0=[Q_{i},Q_{j}]$ but anti-commute
with all the operators of another sets, \emph{i.e.}
$P_{i}Q_{j}+Q_{j}P_{i}=0$. They also satisfy
$P_{i}^{2}=1=Q_{j}^{2},~\forall i,j$ and $[H, P_i(Q_j)] =0$. These
commutation relations imply the existence of doubly degenerate states,
combining the advantages of previously introduced DFS \cite{Roos2004,
  Haff2005, Wine2005}. Indeed, a state corresponding to an even value
of the projection along the $y$-axis along one row is converted into
an odd parity state by the action of $Q_{j}$ operator that commutes
with the Hamiltonian.  Finally, the numerical diagonalization of the
Hamiltonian~(\ref{eq:infinite}) shows that the two ground states are
separated from the rest of the spectrum by a gap $\Delta$ which
depends weakly on the size of the system.

We now discuss the effect of a generic local noise described by the
operator
\begin{equation}
\mathcal{N}=\sum_{i,j}^{N^{2}}(b_{i,j}^{x}\sigma
_{i,j}^{x}+b_{i,j}^{y}\sigma _{i,j}^{y}+b_{i,j}^{z}\sigma _{i,j}^{z}),
\label{noise}
\end{equation}
where the $b_{i,j}^{x,y,z}$ are arbitrary time-dependent coefficients.
Note that no other hypothesis is needed concerning the noise besides
its local character: this noise can be due to energy fluctuations
caused by random fields acting on the atomic system (described by
$\hat{\sigma}_{z}$ operators) as well as spontaneous emission
(described by $\hat{\sigma}_{-}$ operators). Equation (\ref{noise})
should be added to the model Hamiltonian (\ref{eq:infinite}).
Individual terms of (\ref{noise}) do not commute with some of the
symmetries, $P_{i}$ and $Q_{j}$, partially lifting the ground state
degeneracy. When only $k<N$ such terms are present $2N-2k$ symmetries
remain intact ensuring the degeneracy of the ground states levels.
Thus, decoherence is due to the simultaneous action of $N$ noise
terms, in a processes taking place in the $N^{th}$ order perturbation
theory. The small parameter controlling the perturbation development
is $\max (b_{i,j}^{x,y,z})/\Delta $. Decoherence rate acquires an
additional small factor of $\left( \max (b_{i,j}^{x,y,z})/\Delta
\right) ^{N-1}$ when $N$ noise terms act simultaneously. Note that the
decoherence rate of a system consisting of $N^{2}$ interacting
particles usually increases with the number of particles, a behavior
which is in clear contrast to the one observed here. The resulting
decoherence rate is thus
\begin{equation}
\Gamma _{p}=\alpha _{N}\Gamma _{0}\left( \max (b_{i,j}^{x,y,z})/\Delta
\right) ^{N-1},  \label{Gamma_Protected}
\end{equation}
where $\alpha _{N}$ is a numerical constant of the order of the unity.
For this protection to be efficient the gap $\Delta $ should remain
large as $N$ increases.

The results of the exact numerical diagonalization of the long range
Hamiltonian (\ref{eq:infinite}) with $J_{x}=J_{y}$ reported in
Tab.~\ref{tab:gapvalues} show that this is indeed the case. This
property of~(\ref{eq:infinite}) is a clear advantage over short range
models considered in
\cite{Ioffe2002,Doucot2005,Dorier2005,Zoller2006}. It implies that the
suppression of the noise is more efficient in the present system.
Indeed, numerical calculations show that the static fields
$b_{i,j}^{x,y,z}$ randomly distributed in the interval $(-0.1,0.1)J$
result in a very small splitting of the degenerate doublet $\delta
E\sim 10^{-4}-10^{-5}J$ , two orders of magnitude better than that of
\cite{Doucot2005} in agreement with the estimates above.

\begin{figure}[h]
\centering
\includegraphics[width=.5\columnwidth]{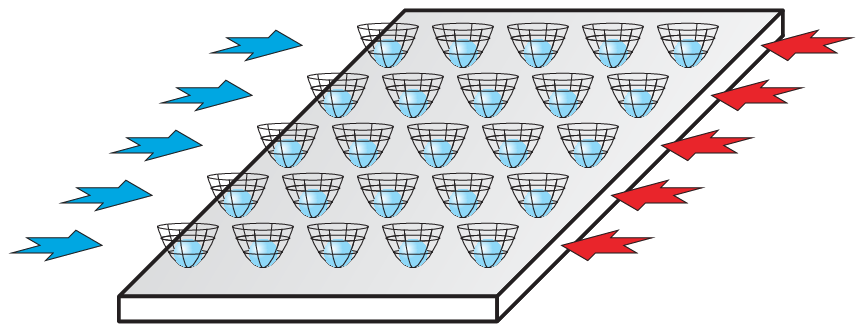}%
\includegraphics[width=.5\columnwidth]{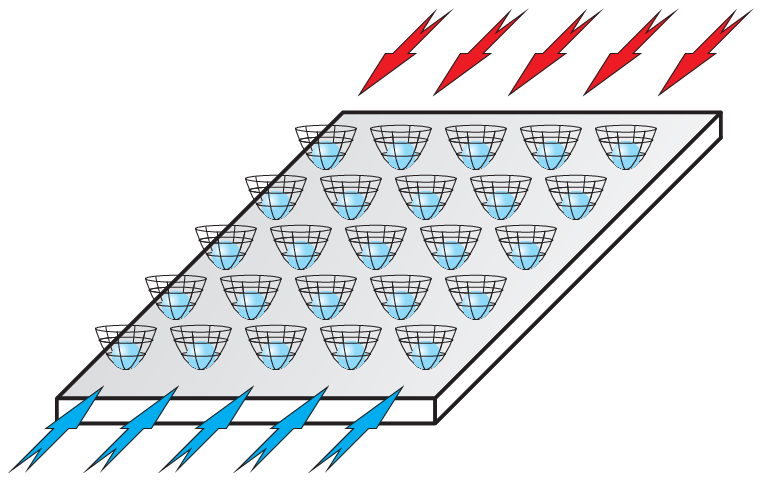}
\caption{Two laser beams of slightly different frequencies (blue and red)
  produce virtual processes that change the state of one ion and
  create (annihilate) a phonon mode of $N\times N$ lattice. Each of
  these laser fields alone does not produce a real ion transition
  (with or without a phonon mode emission or absorption). The sum of
  their frequencies is exactly equal to the energy needed to change
  the state of two ions simultaneously
which is described by the $\protect\sigma _{i,j}^{x,y}\protect\sigma %
_{i,k}^{x,y}$ in the effective Hamiltonian. If the phonon mode
involved in this process corresponds to global translational motion of
all ions (center-of-mass mode), the effective pair-wise interaction
does not depend on the distance between two ions. It can be restricted
to the desired row (column) form if only a subset of ions is
illuminated at a given time. The effect of the sequential application
of the beams to rows and columns is described by the effective
Hamiltonian \protect\ref{eq:infinite}.}
\label{fig:microtrap}
\end{figure}

We now discuss possible implementations of the mathematical model
(\ref{eq:infinite}) in a physical system. As was shown previously,
short range Hamiltonians with topologically protected doublets can be
realized by Josephson junction arrays~\cite{Ioffe2002,Doucot2005}.
Although promising, this approach is difficult due to the required
high degree of similarity of a large number of nanoscale junctions.
Very recently it was shown that it is possible to realize these
Hamiltonians by polar molecules in optical lattices, \cite{Zoller2006}
but this is also technologically difficult. The long range Hamiltonian
(\ref{eq:infinite}) has the advantage, besides the stronger protection
described above, that it can be realized in a trapped ion system with
present day technology using laser induced coupling proposed in
\cite{Sorensen1999} (see Fig.~\ref{fig:microtrap}) . This type of
interaction has already been used to create multiparticle entangled
states in \cite{Wineland2000}.

\begin{figure}[h]
  \centering
\begin{minipage}{.49\columnwidth}
  \includegraphics[width=.75\columnwidth]{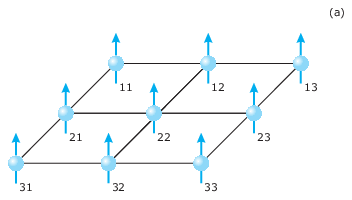}
  \end{minipage}
  \begin{minipage}{.49\columnwidth}
    \includegraphics[width=.95\columnwidth]{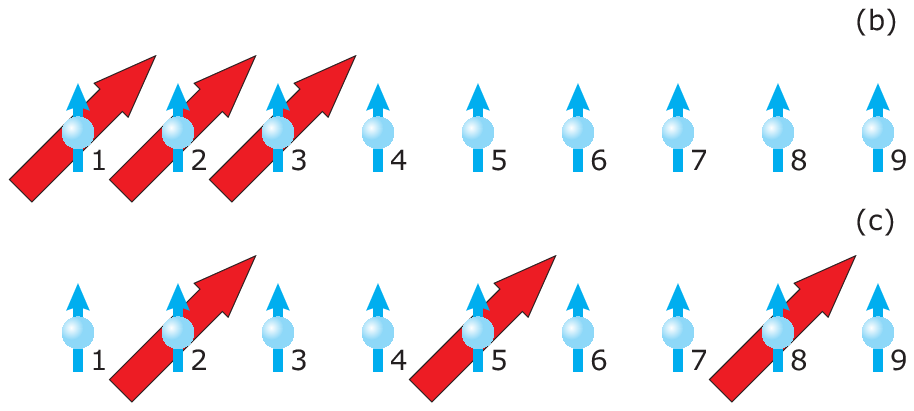}
  \end{minipage}
  \caption{An array of $3\times 3$ implemented by a linear trap with 9 ions.
    To generate the interaction in row 1 (insert a)\ one applies laser
    light to ions 1, 2, 3 (insert b) while column 2 interaction
    requires light on ions 2, 5 and 8. (insert c). The advantage of
    this scheme is that it does not require micro trap fabrication.
    However, it becomes difficult to implement for large arrays
    ($N>3$) because the distance between the ions in the center of the
    array gets smaller making their individual addressing difficult.}
\label{fig:squaretostring}
\end{figure}

In this approach the physical system consists of $N^{2}$ ions that can
either be arranged in a two dimensional array of microtraps, see
Fig.~\ref {fig:microtrap}, or in a linear trap, see
Fig.~\ref{fig:squaretostring}. In both situations, strings of ions are
illuminated by laser fields tuned close to a particular resonance
frequency of individual ions\cite{Sorensen1999,sorensen:pra,james}. In
these conditions the relevant degrees of freedom of each ion can be
represented by a pseudospin $S=1/2$ variable.  As discussed below one
can choose the laser fields so to generate an effective Hamiltonian of
these pseudospins coupling ions two-by-two. Such a pseudospin can be
implemented using an electronic ground state and an excited level
directly coupled by a single photon transition~\cite{Roos1999}, in
this case spontaneous emission will not bring the system out of the
pseudo spin space. Furthermore, for long lived states of alkaline
earth ions\footnote{$ D_{3/2,5/2}$ multiplets have a typical lifetime
  of 1~s}, the corresponding weak metastability induces a small noise
term of the form~(\ref{noise}), 

which is efficiently treated by the
incipient protection mechanism of Hamiltonian~(\ref{eq:infinite}).

We now sketch the derivation of the effective Hamiltonian
(\ref{eq:infinite} ) and estimate realistic constraints on its
parameters. Each ion is in one of two quantum states with an energy
difference $\omega _{eg}$, the interaction between ions produce a
collective vibrational mode of frequency $\nu $ corresponding to the
global displacement of all ions. The ions are subjected to two laser
fields of frequencies $\omega _{1,2}=\omega _{eg}\pm (\nu +\delta )$
and Rabi frequency $\Omega $. In the interacting representation the
physical Hamiltonian describing the ion system in the $ \omega =\omega
_{eg}+(\nu +\delta )$ laser field is \cite{sorensen:pra,james} :
\begin{equation}
H_{int}=\Omega J_{+}e^{-i(\nu +\delta )t}+\frac{i\eta \Omega }{\sqrt{2}}
(a^{\dag }+ae^{-2i\nu t})J_{+}e^{-i\delta t}+h.c.
\end{equation}
where $\eta $ is the Lamb-Dicke parameter defined as $(\hbar
^{2}k_{j}^{2}/2M\hbar \nu )^{1/2}$ and $a$ and $a^{\dag }$ are the
annihilation and creation operators of the phonon mode. Provided that $%
\Omega \ll \nu $ and $\eta \ll 1$ the main contribution to the
effective Hamiltonian comes from the terms oscillating with frequency
$\delta $. For small interaction constant $\eta \Omega \ll \delta $
these terms can be treated in the perturbation theory resulting in the
desired $\chi (\mathbf{\ J\cdot n})^{2}$ term of the effective
interaction where $\mathbf{n}$ is a unit vector whose direction in
$xy$ plane is controlled by the relative phases of the laser
fields\cite{Sorensen1999}. Generally, a stronger interaction might
lead to the excitation of the phonon mode. This process can be
suppressed\cite{sorensen:pra} by choosing the time of the interaction
$\tau $ in such a way that the phonon system returns to its initial
state, $\tau \delta =2 \pi K$. In this case the effective interaction
remains $\chi ( \mathbf{J\cdot n})^{2}$ with $\chi =\frac{\eta
  ^{2}\Omega ^{2}}{\delta }$.

Because the interaction occurs only between the ions illuminated at
one time, the spatial form of the physical array does not need to be
directly related to the column/row form of the effective Hamiltonian
(\ref{eq:infinite}). Thus, as we illustrate in
Fig.~\ref{fig:squaretostring}, a possible configuration rendering the
implementation of the Hamiltonian easier in existing systems, consists
of using a linear array of ions.
\begin{table}[h]
  \centering
  \begin{tabular}{m{1.5cm}>{$}m{1.5cm}<{$}>{$}m{1.5cm}<{$}>{$}m{1.5cm}<{$}>{$}m{1.5cm}<{$}} 
    & 2 \times 2 & 3 \times 3 & 4 \times 4 & 5 \times 5 \\ \hline\hline
    SRI & 0.84 ~J_x & 0.58 ~ J_x & 0.32~ J_x & 0.20~ J_x \\
    LRI & 0.84 ~J_x & 0.96 ~J_x & 0.92 ~J_x & 0.80 ~J_x%
  \end{tabular}
  \caption{Gap as a function of the array size for short range interaction
    (SRI) and long range interaction (LRI)}
  \label{tab:gapvalues}
\end{table}

This model also assumes that the lasers interact only with mode of
frequency $\nu $. This becomes difficult to achieve in large systems
where the number of modes is large. Small coupling to these modes will
induce a position dependent coupling in the effective Hamiltonian
modifying only slightly the gap in the spectrum. Large coupling to
these modes would lead to their excitation which is much more
dangerous. To avoid this process we need to ensure that $\max (\delta
,\Omega )<\nu -\nu _{1}$ where $\nu _{1}$ is the frequency of the
closest mode. In the case of a string of ions, $\nu -\nu _{1} \approx
(1-\sqrt{3})\nu$ and is roughly independent of the number of ions
\cite{steane}, for a fixed trapping potential. Thus, in order to
maximize interaction, we need to adjust the parameters so that $\nu
-\nu _{1}$ is maximal while keeping the ions at distances sufficiently
far apart so that they can be addressed individually. This leads to
conditions on the translational mode frequency, which can be satisfied
experimentally even for chains of 8 ions \cite{blatt_nature}. Using
the results of~\cite{steane,morigi04} for the phonon spectrum of a
single ion chain and assuming a fixed inter-ion distance, we find that
$\nu -\nu _{1}\approx \frac{1}{N^{2}}\times 10$~MHz, which implies
that such implementation is feasible for $N \leq 3$.

To implement larger arrays one can use a two dimensional configuration
of surface traps of a few meV depth, similar to those used in
\cite{surfacetrap}. When the traps are at a distance $a_{min}$ from
each other the Coulomb repulsion becomes of the same order of
magnitude as the trapping potential. In this case, one can achieve a
frequency of the translational modes of the order of $\nu =10$~MHz
\cite{surfacetrap} and our numerical results show that for a $5\times
5$ square array configurations $\nu -\nu _{1}\approx 0.1\nu $.
Coupling between rows and columns is then avoided by applying lasers
that alternate between lines and columns. The resulting Hamiltonian is
the sum of the terms in (\ref{eq:infinite}) provided that the duration
of each pulse satisfies $\tau J_{x,y}\lesssim 1$. We summarize the
experimental values and resulting couplings in
Tab.~\ref{tab:parameters}.

\begin{table}[h]
  \centering
  \begin{tabular}{>{$}m{2cm}<{$}>{$}m{2cm}<{$}>{$}m{2cm}<{$}>{$}m{2cm}<{$}}
  & 4 \text{ ions} & 9 \text{ ions} & 5 \times 5 \text{ ions} \\ \hline\hline
  \Omega (Hz) & 10^5 & 10^5 & 10^6 \\
  \delta (Hz) & 10^4 & 10^4 & 10^5 \\
  J_{x,y} (Hz) & 10^4 & 10^4 & 10^5 \\
  &  &  &
\end{tabular}
\caption{Parameters and induced coupling for the various arrays, $\protect
  \nu =1$~MHz for 4 and 9 ions and 10~MHz pour $5\times 5$ ions, $\protect\eta 
  =0.1$ in both cases, $J_{x,y}=\protect\eta ^{2}\Omega ^{2}/\protect\delta $,
  $K=1$.}
\label{tab:parameters}
\end{table}

The strength of the induced interaction, $J_{x},J_{y}$ and thus the
gap in the spectrum $\Delta $ of the Hamiltonian (\ref{eq:infinite})
is limited by the condition that the laser do not lead to the
excitation of spurious phonon modes. Thus, to achieve a maximal
interaction strength one needs a vibrational spectrum with the largest
possible gap between the global translational mode and the rest of the
spectrum. Another constraint comes from the condition that the
distance from the ions is sufficiently large to allow for their
individual addressing by laser beams. This gap is very sensitive to
the geometry of the lattice. We have identified the two most promising
candidates mentioned above: small ($4$ and $9$ ions) one dimensional
(Fig.~\ref{fig:squaretostring}) and larger ($5\times 5$) highly
symmetric two-dimensional structures (Fig.~\ref{fig:microtrap}).

We now estimate the effective decoherence rates which can be obtained
in these three systems in realistic conditions. Assuming that the
allowed minimal separation between ions is $2\mu$m we get
(Tab.~\ref{tab:parameters}) that the induced couplings $J_{x,y}$ of
the effective Hamiltonian are $10^{4}$~Hz for the linear array and
$10^{5}$~Hz for the square array. As shown in
Tab.~\ref{tab:gapvalues}, the gap is of the same order of magnitude as
the coupling coefficient in the isotropic case, $J_{x}=J_{y}$.

The effective decoherence rate (\ref{Gamma_Protected}) is determined
by the largest noise term and by the single ion decoherence rate. In
an ideal system where the only origin for noise would be the
individual ion decoherence ($b_{i,j}^{x,y,z}\approx \Gamma _{0}$), the
effective lifetime would be astronomical (assuming a minimal
separation between ions
$a_
{min}=2\mu
m$ we get $10^{25}$s for a 5$\times $ 5 array). In a more realistic
situation, one must take into account noises induced by the ions
manipulation with laser light: the dominant noise turns out to come
from laser frequencies noise $\delta f$.  Frequencies drift with time
resulting in a $H=\hbar \delta f\sum_{i,j} \sigma _{i,j}^{z}$ term in the
effective Hamiltonian,. Though one can suppress this noise down to the
level of some Hz \cite{wineland:nist} we shall use a more realistic
value of 500~Hz for our estimates summarized in
Tab.~\ref{tab:Gammaeff}.

\begin{table}[h]
  \centering
\begin{tabular}{>{$}m{2cm}<{$}>{$}m{2cm}<{$}>{$}m{2cm}<{$}>{$}m{2cm}<{$}}
  & 4 \text{ ions} & 9 \text{ ions} & 5 \times 5 \text{ ions} \\ \hline\hline
\Gamma_{eff} (Hz) & 1.5 \cdot 10^{-3} & 7.5 \cdot 10^{-5} & 1.9 \cdot
10^{-11} \\
\tau (s) & 6.6 \cdot 10^{2}  & 1.3 \cdot 10^{4} & 5.3 \cdot 10^{10}
\end{tabular}
\caption{Effective decoherence rates and topological qubit lifetimes.}
\label{tab:Gammaeff}
\end{table}

Global state initialization can be achieved by first imposing a large
effective field along the $x$ direction, yielding a large additional
term proportional to $H_{\mathrm{ext}}=\pm \sum
\hat{\sigma}_{i,j}^{x}$ in~(\ref{eq:infinite}). This leads to the
global state of the array $\ket 0 =\prod \ket 0 _{ij}$ or $\ket 1
=\prod \ket 1 _{ij}$ depending on the overall sign in
$H_{\mathrm{ext}}$. These states satisfy $Q_{j}\ket 0 =\ket 0 $ and
$Q_{j}\ket 1 =-\ket 1 $ (if $N$ is odd). When $H_{ext}$ is switched
off adiabatically (i.e on a time scale much longer than $1/\Delta $),
the final Hamiltonian becomes~(\ref{eq:infinite}). At all stages of
this evolution the Hamiltonian commutes with $Q_{j}$ operators.  Thus,
the two initial unprotected states $\ket 0$ and $\ket 1$ evolve into
the protected states with the same quantum numbers. Note that $P_{i}$
operators commute only with the final Hamiltonian, and the initial
states are not eigenstates for them. The final state can be
reconstructed using standard methods \cite{blatt_nature} measuring the
state of individual ions in the $\sigma ^{x}$ or $\sigma ^{y}$ basis,
after the protecting interaction~(\ref{eq:infinite}) has been switched
off.

We have shown that it is possible to use the flexibility of ion traps
to form long living quantum states topologically protected from
decoherence, even small 4 ion strings being expected to give
relaxation times one order of magnitude longer than currently achieved
(Tab.~\ref{tab:Gammaeff}).  It is quite likely that similar techniques
can be used for experimental implementation of more exotic quantum
states in controlled atomic systems, for instance the states described
by the anyon statistics that would make possible to implement a full
quantum computation in the protected subspace.


\end{document}